\documentclass[10pt,conference]{IEEEtran} 
\IEEEoverridecommandlockouts
\usepackage{cite}
\usepackage{multirow}
\usepackage{comment}
\usepackage{amsmath,amssymb,amsfonts}
\usepackage{algorithmic}
\usepackage{graphicx}
\usepackage{textcomp}
\usepackage[table,xcdraw]{xcolor}
\def\BibTeX{{\rm B\kern-.05em{\sc i\kern-.025em b}\kern-.08em
    T\kern-.1667em\lower.7ex\hbox{E}\kern-.125emX}}
\begin{document}

\title{Aerial Base Stations: Practical Considerations for Power Consumption and Service Time}

\author{
 \IEEEauthorblockN{Siva Satya Sri Ganesh Seeram\IEEEauthorrefmark{1},
Shuai Zhang\IEEEauthorrefmark{1}, Mustafa Ozger\IEEEauthorrefmark{1}, Andre Grabs\IEEEauthorrefmark{2},  Jaroslav Holis\IEEEauthorrefmark{3}, Cicek Cavdar\IEEEauthorrefmark{1}}
\IEEEauthorblockA{\IEEEauthorrefmark{1} KTH Royal Institute of Technology, Sweden, Email: \{sssgse, shuai2, ozger, cavdar\}@kth.se \\
\IEEEauthorrefmark{2} Airbus, Germany, Email: andre.grabs@airbus.com
\\
\IEEEauthorrefmark{3} Deutsche Telekom, Germany, Email: jaroslav.holis@t-mobile.cz
}
}

\maketitle

\begin{abstract}

Aerial base stations (ABSs) have emerged as a promising solution to meet the high traffic demands of future wireless networks. Nevertheless, their practical implementation requires efficient utilization of limited payload and onboard energy. Understanding the power consumption streams, such as mechanical and communication power, and their relationship to the payload is crucial for analyzing its feasibility. Specifically, we focus on rotary-wing drones (RWDs), fixed-wing drones (FWDs), and high-altitude platforms (HAPs), analyzing their energy consumption models and key performance metrics such as power consumption, energy harvested-to-consumption ratio, and service time with varying wingspans, battery capacities, and regions. Our findings indicate that FWDs have longer service times and HAPs have energy harvested-to-consumption ratios greater than one, indicating theoretically infinite service time, especially when deployed in near-equator regions or have a large wingspan. Additionally, we investigate the case study of RWD-BS deployment, assessing aerial network dimensioning aspects such as ABS coverage radius based on altitude, environment, and frequency of operation. Our findings provide valuable insights for researchers and telecom operators, facilitating effective cost planning by determining the number of ABSs and backup batteries required for uninterrupted operations.
    
\end{abstract}

\begin{IEEEkeywords}
    Unmanned aerial vehicles, aerial base stations, energy harvesting, power consumption, service time, aerial network.
\end{IEEEkeywords}

\setlength{\textfloatsep}{0.3pt}
\vspace{-4mm}
\section{Introduction}
\vspace{-2mm}
Flying base stations have been proposed as a candidate solution to provide cellular connectivity to ground users, especially in inaccessible areas, or to boost the capacity of existing terrestrial networks in hot spots. This is achieved by installing a base station (BS) on unmanned aerial vehicles (UAV), also known as aerial base station (ABS). Despite the widespread use of ABSs, their practical implementation is often overlooked. For their practical deployment, the characteristics and capabilities of different types of ABSs such as rotary-wing drones (RWDs), fixed-wing drones (FWDs), and fixed-wing type high-altitude platforms (HAPs)~\cite{b1,b9} need to be carefully analyzed. One of the primary limitations is their size, weight, and power (SWAP) characteristics. 
Extensive research has been conducted on optimizing the energy consumption of ABSs through trajectory planning, resource allocation, and management techniques, focusing on mechanical and communication power consumptions~\cite{bb1}. However, a comprehensive understanding of the various power consumption streams is also crucial for developing a holistic perspective on this issue. Unfortunately, most existing studies often overlook the communications power, assuming it to be negligible compared to the mechanical power consumption~\cite{bb2}, thereby neglecting the associated communications payload power. Our study seeks to address this gap by providing an analysis of power consumption and assessing the validity of such key assumptions in this area.
A primary focus of sixth-generation (6G) cellular networks is the introduction of novel wireless technologies and innovative network architectures including airborne vehicles. Moreover, these architectures can help in achieving ubiquitous coverage. 
Hence, the practical deployment of ABSs raises questions about the compatibility and feasibility of various types of BSs, such as macrocell, microcell, and picocell, on UAVs. Feasibility is defined as the ability to deploy the BS on the aerial platform in terms of payload, while compatibility refers to whether the feasible BS can satisfy the coverage requirements when deployed on a particular aerial platform. Terrestrial BSs vary in weight, transmission range, capacity, and coverage areas, requiring a delicate balance to meet SWAP constraints to use as non-terrestrial BSs. Despite the practical deployment, one important factor to consider is the power consumption impact stemming from each type of BS~\cite{b3,b12,b8} mostly dominated by the payload impact. By analyzing this impact on the total power consumption and capacity of each BS, one can determine the most suitable deployment on UAVs specific to use cases and optimize their performance for communication purposes. Furthermore, due to the payload limitations, certain BS functionalities should be placed on the ground to decrease the weight of the ABSs. 

The energy harvesting capabilities of ABSs may ensure their sustainability. Solar energy harvesting is currently the most promising one, particularly for HAPs due to their increased exposure to the solar flux at higher altitudes. However, the potential benefits and drawbacks of utilizing solar energy for energy harvesting and its robustness and overall impact on the sustainability of ABS should be thoroughly explored.
This study analyzes the power consumption, energy harvested-to-consumption ratio, and service time of ABSs. The energy harvested-to-consumption ratio is defined as the ratio of the total energy harvested by the solar panel to the total energy consumption in a day, which indicates the robustness of the ABSs. Service time refers to the duration in which the ABS remains operational and serves ground users. However, this analysis excludes take-off and landing power consumption and associated flight time. Power consumption values particularly change with different payloads, which comprise the weight of the BS, battery, and solar panels installed on the ABS. An important research question is whether the energy harvested from solar panels deployed on these different ABSs is sufficient to power them. Additionally, it is crucial to investigate how the added weight and power consumption of the BS affect the ABS's total flight time. This entails analyzing the ABS's capacity and coverage requirements to serve a particular area for a specific duration. Moreover, the deployment cost for such scenarios involves determining the number of ABSs and backup batteries required to ensure continuous operation. Their dependency on the charging power is another aspect to understand the potential of ABS deployments. This study is the first attempt to investigate the feasibility and practicality of ABS deployment, which has not been studied from SWAP perspective. The main contributions of this study are:
\vspace{-1mm}
\begin{itemize}
    \item To the extent of our knowledge, our study is the first and only work that compiles the practical challenges and limitations of different aerial platforms and BSs and analyzes various power consumption streams.
    \item Our research evaluates the feasibility and compatibility of deploying different ABSs. 
    \item We summarize the power consumption models of different ABSs and demonstrate how solar energy harvesting affects their service time.
    \item Extensive simulations are performed to investigate the aspects such as coverage radius, number of ABSs, and backup batteries required in a case study.

\end{itemize}

The remainder of the paper is organized as follows. Section~\ref{sec:ABS power} outlines the power streams and solar energy harvesting models for ABSs. Section~\ref{sec:Serv time} investigates the service time associated with different ABSs and explores the feasibility of diverse ABS configurations. The network dimensioning is then examined through a case study focusing on RWD-ABS in Section~\ref{sec:cov ana}, followed by the presentation and analysis of the case study results in Section~\ref{sec:res}. Finally, the paper concludes with a summary of the findings in Section~\ref{sec:conclu}.




\vspace{-1mm}
\section{Power Consumption Breakdown Analysis}
\vspace{-1mm}
\label{sec:ABS power}

\begin{figure}[b!]
    \centering
    \includegraphics[width=0.9\linewidth]{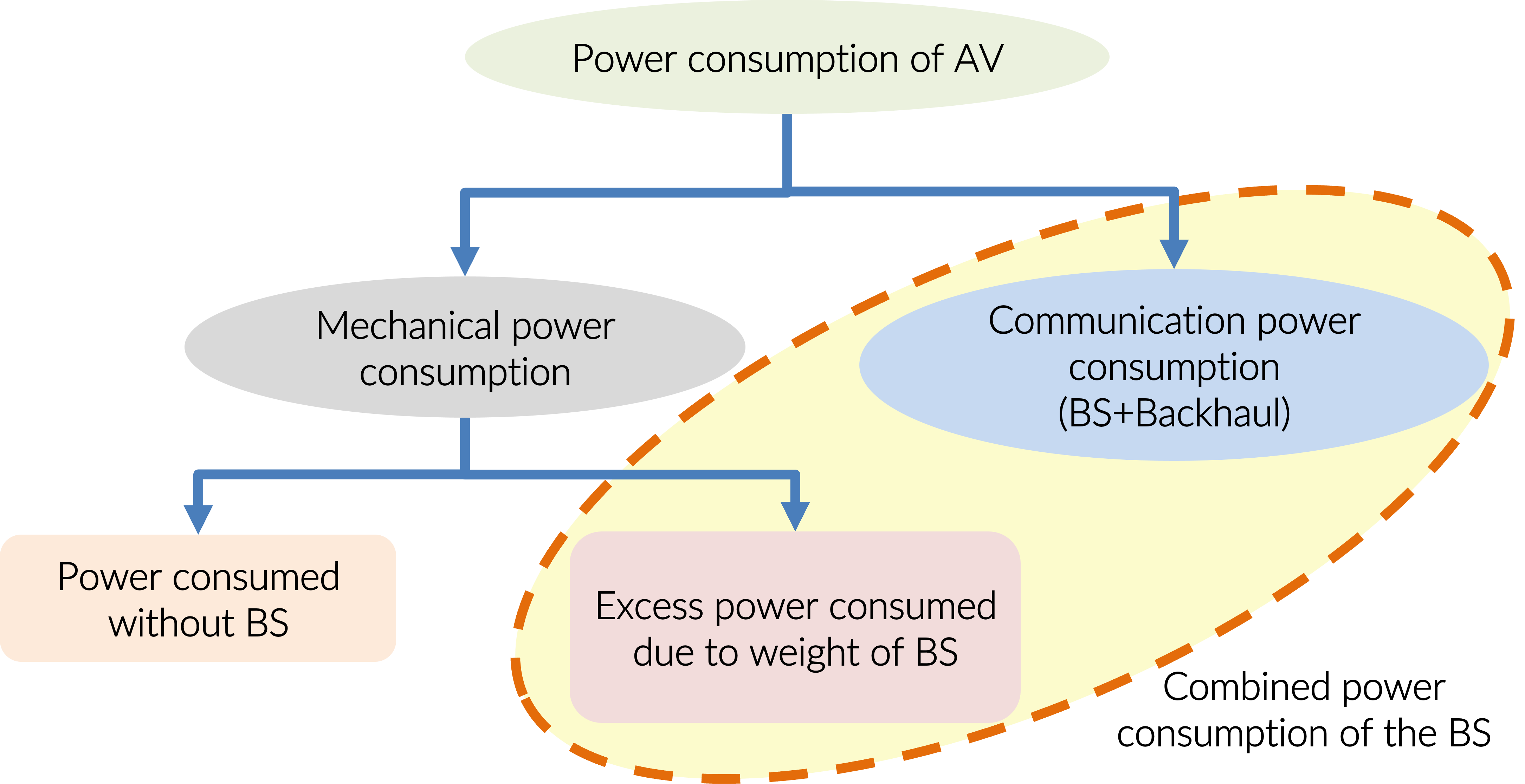}
    \vspace{-3mm}
    \caption{Different power streams in ABSs.}
    \label{fig:powerstream}
\end{figure} 

The service time of an ABS is affected by the power consumption, which stems from mechanical and communication system operations as seen in Figure~\ref{fig:powerstream}. The former refers to the power required by the ABS to remain aloft assuming full utilization of payload. The latter refers to the power utilized to operate the BS installed on the ABS. Another key aspect is that the communication power consumption and excess power consumed due to BS weight are interrelated. When there is no BS deployed on the UAV, these two power streams do not exist. Hence, we term these two power streams as the combined power consumption of the BS. Moreover, in addition to comprehending the power budget, understanding the total mass budget of the ABS is crucial. This encompasses the sum of the structural platform mass and the payload. In the case of RWDs, the payload refers to the mass of the BS, solar panels, and batteries, while for others, it refers to the BS payload only.
\vspace{-6mm}
\subsection{Power Consumption Model for Rotary-Wing Drone}
\vspace{-1mm}
The  quantitative estimate of power consumption for a hovering configuration of the RWD~\cite{b10} is given as
\vspace{-2mm}
\begin{equation}
\label{eq:rwd1}
    P_{RW}=\frac{\delta}{8}\rho s A_D \Omega^3 R^3 + (1+k)\frac{W^{3/2}}{\sqrt{2\rho A_D}},
\vspace{-1mm}
\end{equation}
where $\delta$ is profile drag co-efficient, $k$ is incremental correction factor, $W$ is the total weight of the RWD, $\rho$ is the air density, $s$ is rotor solidity, $A_D$ is rotor disc area, $\Omega$ is the blade angular velocity, $R$ is the rotor radius. Our analysis focuses solely on the power required for hovering the RWD. 
\vspace{-2mm}
\subsection{Power Consumption Model for Fixed-Wing Drone}
\vspace{-1mm}
The propulsion power consumption for maneuvering the FWD given that it follows a circular trajectory~\cite{b11} with radius $r$ and speed $V$ is calculated as
\vspace{-2mm}
\begin{equation}
\label{eq:fwd1}
    P_{FW}(V,r)=(c_1+\frac{c_2}{g^2r^2})V^3+\frac{c_2}{V}   ,
\vspace{-1mm}
\end{equation}
where $g$ is the gravitational acceleration, $c_1$ and $c_2$ are the parameters related to the FWD~\cite{b11} given as follows
\vspace{-2mm}
\begin{equation}
\label{eq:fwd2}
    c_1=\frac{1}{2}\rho C_{D_0}A_w, \mathrm{ and }\; c_2=\frac{2W^2}{\pi e_0 \mathrm{A_R}\rho A_w},
\vspace{-1mm}
\end{equation}
where $C_{D_0}$ is the zero-lift drag coefficient, $A_w$ is the wing area, $e_0$ is Oswald efficiency, $\mathrm{A_R}$ is the aspect ratio of the wing, and $W$ is the weight of the aircraft.
\vspace{-2mm}
\subsection{Power Consumption Model for Fixed-Wing HAP}
\vspace{-1mm}
 In fixed-wing type HAPs, mechanical power consumption is the summation of the power consumption of propulsion and avionics. Avionics power consumption is calculated using the mass-to-power ratio given as $6$ W/kg, and the weight of the avionic sub-system is considered as $22$ kg~\cite{b9}. However, propulsion power consumption is lower at higher altitudes due to lower air density. The propulsion power consumption of the HAPs~\cite{b9} is given as follows
 \vspace{-2mm}
\begin{equation}
\label{eq:HAP1}
    P_{HAP}=\frac{C_D}{\eta_{p}C_L^{3/2}}\sqrt{\frac{2W^3}{\rho A_w}},
\vspace{-1mm}
\end{equation}
where $C_D$ is the drag coefficient, $C_L=2W/(\rho V^2A_w)$ is the lift coefficient, $\eta_{p}$ is the propeller efficiency. 
\vspace{-2mm}
\subsection{Base station Power Consumption Model}
\vspace{-1mm}
The selection of a suitable BS to deploy on UAVs depends on various factors such as the intended use case. The BS options include picocells, microcells, and macrocells, among others, each with different power consumption characteristics. The power consumption for different baseband functions is given in terms of computational complexity measured in giga operations per second (GOPS) and intrinsic efficiency is used to compute the power consumption of each baseband function and power amplifier~\cite{b3}. A comprehensive list of power consumption values for each type BS for a given configuration is presented in~\cite{b12} and is given by
\vspace{-1mm}
\begin{equation}
\label{eq:BS1}
    P_{BS}=P_{BB}+P_{RF}+P_{PA}+P_{OH},  
\end{equation}
where different power consumption streams are $P_{BB}$ related with the baseband functions, $P_{RF}$ the radio frequency (RF) power, $P_{PA}$ the power amplifier, and $P_{OH}$ the overhead power. For full BS deployment case, we use~\eqref{eq:BS1}, however, for split case deployment, we ignore $P_{BB}$ since the baseband unit will be off-loaded to the ground.
\vspace{-2mm}
\subsection{Backhaul Power Consumption Model}
\vspace{-1mm}
The backhaul link between the ground station and the ABS is considered the most efficient, as it is not limited by SWAP constraints as compared to the service link. The limitations on the service link can be offset by offloading some BS functions to the ground station. In general, backhaul weight, size, and power consumption are assumed to be less than $10$\% of those for the service link. There are several potential backhaul technologies like free space optics (FSO) and millimeter-Wave (mmWave) between ground-ABS, ABS-ABS, and ABS-satellite communications. 
\vspace{-2mm}
\subsection{Energy Harvesting Model}
\vspace{-1mm}
Energy harvesting may extend ABSs' operational time in the air. The total energy harvested by the solar panel for the whole day is calculated as
\vspace{-1mm}
\begin{equation}
\label{eq:Solar1}
    E_{T}=A_{pv}G_{T}\eta_{pv},  
\vspace{-1mm}
\end{equation}
where $A_{pv}$ is the total area of installation of the photovoltaic system, $\eta_{pv}$ is the efficiency of the photovoltaic cell, $G_{T}$ is the total solar irradiance per m$^2$ per day~\cite{b9}. 

From the analysis of different available solar technologies, throughout the study, we chose thin-film Gallium Arsenide (GaAs) solar cells with an efficiency of $37.75$\% and areal density of $114$ g$\cdot$m$^{-2}$~\cite{b13}. The average $G_{T}$ is a critical parameter influenced by many factors such as latitude, altitude, and day of the year. To investigate the impact of solar radiation, we examine two extreme scenarios: Enugu, a Nigerian city located at $6.4^\circ$  latitude, which is close to the equatorial region, and York, a British city located at $53.9^\circ$  latitude, which is far from the equator. We further select December 21st, which has the shortest daylight period across the year in the northern hemisphere. The $G_{T}$ values for Enugu and York are taken as $10$ kWh/m$^2$/day and $1.5$ kWh/m$^2$/day, respectively~\cite{b9}.  

\vspace{-1mm}
\section{Service Time of Aerial Base Stations}
\vspace{-1mm}
\label{sec:Serv time}
\subsection{Rotary-Wing Drone}
\vspace{-1mm}
The mass budget of RWDs is a significant factor to consider for their service time as observed in~\eqref{eq:rwd1}. For heavy payloads, such as up to $20$ kg~\cite{a1}, the mass budget must be allocated among the battery size, BS weight, and solar panel mass, and the structural mass is assumed as $8$ kg where total take-off mass is the sum of the structural mass and payload. This study considers a picocell and microcell weighing $3$ kg and $12$ kg, respectively. The solar panel area installed on RWD is varied with $0.5$ m$^2$, $1$ m$^2$, and the scenario without solar panels is also evaluated. From the remaining available mass budget, battery mass is varied from $5$ kg to $9$ kg, with $1$ kg increment, and the specific energy ($E_d$) of the battery is changed to \{$50$, $180$, $350$\} Wh/kg~according to battery technologies \cite{b18}. Specific energy is commonly defined as the nominal battery energy per unit mass, which is also known as the gravimetric energy density. To avoid any ambiguity, we have chosen to use the term ``energy density" to specifically refer to this specific energy in our paper. 
\begin{figure}[!b]
\vspace{-1.3mm}
    \centering
    \includegraphics[width={0.92\linewidth}]{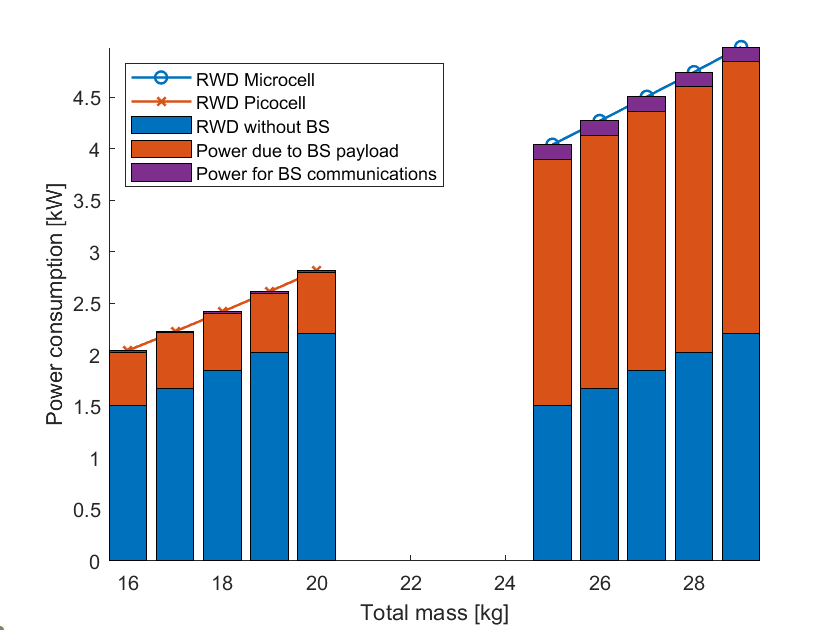}
    \vspace{-4mm}
    \caption{Power consumption vs total mass of RWDs.}
    \label{fig:rwd1}
\end{figure}
The total power consumption ($P_{T,R}$) is calculated as a summation of~\eqref{eq:rwd1}, and~\eqref{eq:BS1}. The parameters used~\cite{b10} to calculate~\eqref{eq:rwd1} are: $\delta=0.012$, $\rho=1.225$ kg$\cdot$m$^{-3}$, $s=0.05$, $R=0.4$ m, $\Omega=300$ rad$\cdot$s$^{-1}$, $k=0.1$.
In Figure~\ref{fig:rwd1}, we can observe the relationship between the power consumption of RWDs and take-off mass for picocell and microcell deployments. The set of bar graphs on the left (right) side of Figure~\ref{fig:rwd1} corresponds to the picocell (microcell) deployment. The variation in each deployment is due to a change in battery mass from $5$ kg to $9$ kg. The power consumption due to BS payload (orange) remains constant for each deployment, as the weights of the picocell and microcell are fixed at $3$ kg and $12$ kg, respectively. Similarly, the power consumption for BS communications (purple) remains constant across all deployments. The increase in the total power consumption of the RWD is solely due to battery mass, as reflected in the power consumed by the RWD without BS (blue). Notably, the communication power consumption is insignificant compared to the total power consumption. However, the orange bar has a more significant contribution to the total power consumption.

\begin{figure}[!tb]
    \centering
    \includegraphics[width={0.9\linewidth}]{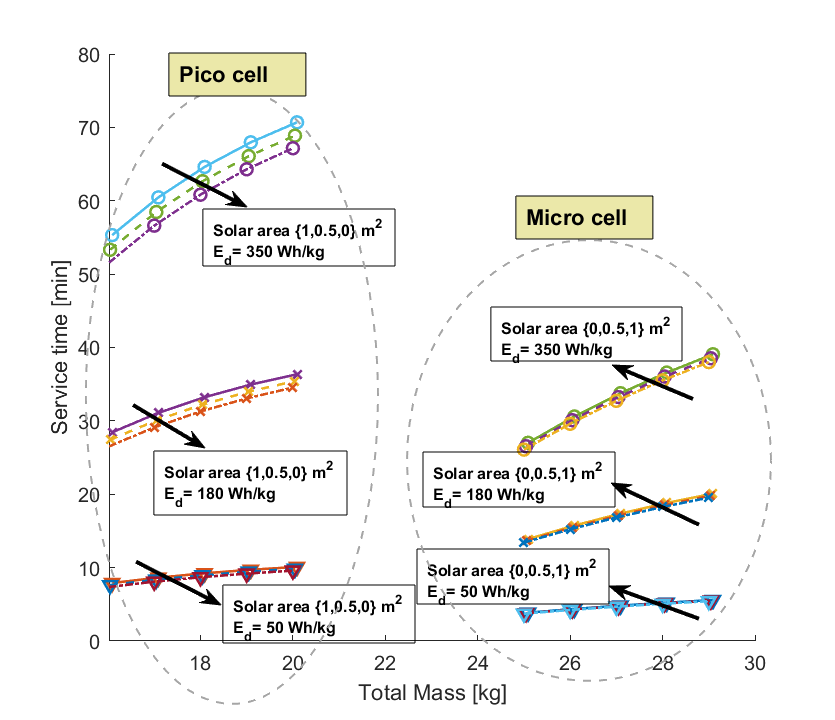}
    \vspace{-4mm}
    \caption{RWD Service time vs total mass for varying $E_d$ and solar area.}
    \label{fig:rwd2}
\end{figure}

Service time of an RWD is calculated using~\eqref{eq:rwd1},~\eqref{eq:BS1} and~\eqref{eq:Solar1}:
\vspace{-1mm}
\begin{equation}
\label{eq:service timeRWD}
    T_{s,R}=\frac{E_dm_{b}}{P_{T,R}-P_{sol}},  
\vspace{-1mm}
\end{equation}
where $m_{b}$ is the battery mass, and $P_{sol}=(E_T/24)$ is the solar harvested power. The parameters that can be varied for studying the service time characteristics are $E_d=\{50, 180, 350\}$ Wh/kg in~\eqref{eq:service timeRWD}, $A_{pv}=\{0, 0.5, 1\}$ m$^2$ in~\eqref{eq:Solar1}. The service time characteristics of RWD-picocell and RWD-microcell deployment are calculated by varying the  energy density and installation of solar panels with different sizes of the area as shown in Figure~\ref{fig:rwd2}. The arrow direction in each set of curves indicates the order of solar panel area, $1$ m$^2$, $0.5$ m$^2$, and $0$ m$^2$ corresponding to the case without the solar panel. It is evident from~\eqref{eq:service timeRWD} and Figure~\ref{fig:rwd2} that the service time is linearly proportional to the $E_d$ and $m_b$.
Our results in Figure~\ref{fig:rwd2} also show that for any picocell deployment, the service time is more than that of the microcell deployment. However, the installation of solar panels on RWD does not have a significant impact on improving the service time. The maximum gain in service time that can be achieved with the installation of solar panels is only up to $5$ minutes for RWD-picocell and a minute for RWD-microcell which is not significant enough considering the cost of installation of solar panels. Note that this solar energy harvesting analysis is done considering the Enugu region. Hence, we may conclude that for York, the highest gain in service time would not even be a minute.
\vspace{-2mm}
\subsection{Fixed-Wing Drone}
\vspace{-1mm}
The propulsion power consumption is calculated using~\eqref{eq:fwd1}. We assume an FWD with a wingspan of $10$ m and a payload capacity of $10$ kg as a baseline. We also consider an FWD
with a wingspan of $5$ m and a payload of $11$ kg \cite{a1}. It is feasible to install pico, micro, and macro BSs weighing less than these payloads, assuming that the mass solar panels are included in the FWD structural mass. The structural mass is assumed to be $44$ kg and $50$ kg, respectively. However, when a picocell is deployed on an FWD flying around $1$ km altitude, it is not compatible due to its short-range coverage of communication. Hence, based on the payload availability, we consider microcell BS deployment for both wingspan configurations.
Finally, we calculate the total power consumption ($P_{T,F}$) via ~\eqref{eq:fwd1} and~\eqref{eq:BS1} by using the input parameters $\rho=1.112$ kg$\cdot$m$^{-3}$, $C_{D_0}=0.0447$, $A_{R}=9.5$, $e_0=0.7548$, $g=9.8$ m$\cdot$s$^{-2}$, $V=20$ m$\cdot$s$^{-1}$, $r=158$ m~\cite{b11}.
Energy harvested-to-consumption ratio~\cite{b9} is calculated using~\eqref{eq:fwd1}, ~\eqref{eq:BS1} and~\eqref{eq:Solar1}, and is given by  
\vspace{-3mm}
\begin{equation}
\label{eq:}
    \psi_F=E_{T}/(24\times P_{T,F}),
\vspace{-2mm}
\end{equation}
$\psi_F$ ratio is used to determine the service time of the aerial platform. When $\psi_F$ is less than $1$, the service time is limited, and it can be computed in hours as $T_{s,F} = 24\psi_F$. Conversely, if $\psi_F$ is greater than or equal to $1$, the theoretical service time becomes infinite due to continuous solar energy harvesting. 
\begin{table}[!t]
\centering
\caption{Estimated power consumption and service times for FWD-ABS.}
\vspace{-3mm}
\label{tab:FWD}
\begin{tabular}{|c|c|c|cc|}
\hline
\multirow{2}{*}{\textbf{Wingspan}} & \multirow{2}{*}{\textbf{Max. Payload}} & \multirow{2}{*}{\begin{tabular}[c]{@{}c@{}}$P_{T,F}$\end{tabular}} & \multicolumn{2}{c|}{$T_{s,F}=24\psi_F$} \\ \cline{4-5} 
 &  &  & \multicolumn{1}{c|}{\textbf{Enugu}} & \textbf{York} \\ \hline
5 m & 11 kg & 1.15 kW & \multicolumn{1}{c|}{8.56 h} & 1.28 h \\ \hline
10 m & 10 kg & 2.39 kW & \multicolumn{1}{c|}{16.47 h} & 2.47 h \\ \hline
\end{tabular}
\end{table}
We investigate the impact on $\psi_F$ by varying the $G_T$ parameter in~\eqref{eq:Solar1}. The resulting $P_{T,F}$ and $T_{s,F}$ (since $\psi_F<1$) with a microcell deployment are tabulated in Table~\ref{tab:FWD}. From the table, we observe that power consumption increases with increasing wingspan. In equatorial regions, the FWD can serve the ground users for at least ($8$ to $16$) h based on the wingspan. However, in regions such as York with less solar radiance, the service time is reduced to ($1$ to $2.5$) h respectively. However, the $10$ m FWD can harvest more energy due to the increased wing area and therefore provides more service time.
\vspace{-2mm}
\subsection{High Altitude Platform}
\vspace{-1mm}
For the HAP analysis, we evaluated four different HAP configurations with varying wingspans and payloads, namely NASA Pathfinder, NASA Centurion, Airbus Zephyr-S, and Phasa-35, in respective order as presented in Table~\ref{tab:HAP}. The structural platform mass of these configurations are ($207$, $592$, $75$, $150$) kg, respectively. We choose to evaluate two categories of HAPs, namely heavy payload HAPs (blue) corresponding to the first two configurations and light payload HAPs (orange) corresponding to the last two configurations. For the last two configurations, due to their limited payload capacities, it is not possible to deploy a full macro BS on the HAP. Instead, we allocate the payload budget within antenna and front-end modules only, with beamforming capability, while the remaining BS components are located on the ground. This allows for a bent pipe architecture with a feeder link operating in the mmWave band and a multibeam service link in the International Mobile Telecommunications (IMT) bands. In the case of the first two configurations, we can deploy macro BS and also can allocate a budget for the solar panel mass within the payload itself, while for the last two configurations, the mass of solar panels is included in the HAP structural platform mass. We analyze two different deployment scenarios for the HAPs, namely the full BS deployment and the split deployment. ``Split" deployment refers to only antenna and front-end modules deployment at the ABS. 

 The total power consumption ($P_{T,H}$) of HAPs is calculated as the summation of~\eqref{eq:HAP1} and~\eqref{eq:BS1} and parameters used are $C_D=0.0071$, $\eta_{p}=0.8$, $\rho=0.08891$ kg$\cdot$m$^{-3}$, $V=20$ m$\cdot$s$^{-1}$~\cite{b9}.
 Similar to FWD, the energy harvested-to-consumption ratio is calculated for HAPs \emph{i.e.,} $\psi_H=E_{T}/(24\times P_{T,H})$. We vary the $G_T$ parameter in~\eqref{eq:Solar1} to analyze the effect on $\psi_H$. we have tabulated $P_{T,H}$ and $\psi_H$ for different regions and BS deployments as in Table~\ref{tab:HAP}. Comparing different deployment scenarios, we observe that $P_{T,H}$ decreases for split deployments due to the removal of $P_{BB}$ in~\eqref{eq:BS1}. The empty entries in the table indicate infeasible BS deployments and HAP configurations. Emphasizing the significance of $\psi_H$, we find that it is greater than $1$ for all configurations, indicating self-sustainability. In the Enugu region, heavy payload HAPs show an increase in the $\psi_H$ value by a magnitude of $2$-$3$ for the split deployment compared to full BS. Light payload HAPs exhibit $\psi_H$ values greater than $7$ and $12$, respectively for split deployment. This demonstrates high harvesting robustness in Enugu. On the other hand, for the case of York with low solar irradiance, we can see that both the BS deployments for a $30$ m wingspan, have $\psi_H$ values with very low margins to $1$, which may not be robust due to the fact that the energy harvesting elements are not 100\% efficient. The practical values would be even lower, making it unreliable to conclude indefinite self-sustainability. However, for a $60$ m wingspan, due to its larger wing area, it has $\psi_H$ greater than $3$. For $25$ m and $35$ m wingspans, a similar analysis can be applied. This implies that a large wingspan HAP or a hydrogen fuel cell backup would be necessary to ensure robustness and self-sustainability in low irradiance regions.
\begin{table}[!t]
\centering
\caption{Estimated power consumption and $\psi_H$ for HAP-ABS.}
\vspace{-2mm}
\label{tab:HAP}
\resizebox{\columnwidth}{!}{%
\begin{tabular}{|c|c|cc|cccc|}
\hline
\multirow{3}{*}{\textbf{\begin{tabular}[c]{@{}c@{}}Wingspan\\ {[}m{]} \end{tabular}}} & \multirow{3}{*}{\textbf{\begin{tabular}[c]{@{}c@{}}Max. Payload\\ {[}kg{]}\end{tabular}}} & \multicolumn{2}{c|}{\multirow{2}{*}{\begin{tabular}[c]{@{}c@{}}$P_{T,H}$ [kW]\end{tabular}}} & \multicolumn{4}{c|}{\textbf{ $\psi_H$}} \\ \cline{5-8} 
 &  & \multicolumn{2}{c|}{} & \multicolumn{2}{c|}{\textbf{Enugu}} & \multicolumn{2}{c|}{\textbf{York}} \\ \cline{3-8} 
 &  & \multicolumn{1}{c|}{\textbf{Full BS}} & \textbf{Split} & \multicolumn{1}{c|}{\textbf{Full BS}} & \multicolumn{1}{c|}{\textbf{Split}} & \multicolumn{1}{c|}{\textbf{Full BS}} & \textbf{Split} \\ \hline \hline
 \rowcolor[HTML]{DAE8FC} 
30 & 45 & \multicolumn{1}{c|}{1.78} & 1.54 & \multicolumn{1}{c|}{7.72} & \multicolumn{1}{c|}{9.6} & \multicolumn{1}{c|}{1.15} & 1.44 \\ \hline
\rowcolor[HTML]{DAE8FC} 
60 & 270 & \multicolumn{1}{c|}{2.68} & 2.44 & \multicolumn{1}{c|}{21} & \multicolumn{1}{c|}{24.24} & \multicolumn{1}{c|}{3.15} & 3.63 \\ \hline \hline
\rowcolor[HTML]{F5DBC0}
25 & 5 & \multicolumn{1}{c|}{-} & 1.45 & \multicolumn{1}{c|}{-} & \multicolumn{1}{c|}{7.1} & \multicolumn{1}{c|}{-} & 1.06 \\ \hline
\rowcolor[HTML]{F5DBC0}
35 & 15 & \multicolumn{1}{c|}{-} & 1.65 & \multicolumn{1}{c|}{-} & \multicolumn{1}{c|}{12.21} & \multicolumn{1}{c|}{-} & 1.83 \\ \hline
\end{tabular}%
}
\end{table}
\vspace{-2mm}
\subsection{Summary of Power Consumption and Feasibility of ABSs.}
\vspace{-1mm}
According to models in Section~\ref{sec:ABS power} and parameter assumptions discussed in the preceding subsections for each ABS, we have computed the power consumption of various ABS configurations, and the results are presented in Table~\ref{tab:power}. For RWD analysis, we considered the configuration with a $5$ kg battery. To provide clarity, we have color-coded the power streams associated with each column in the table in reference to the power streams illustrated in Figure~\ref{fig:powerstream}. These findings underscore the significance of considering the non-negligible communications-associated power consumption in energy optimization efforts, as highlighted in previous studies such as~\cite{bb1,bb2}. A synopsis of the overall feasible and compatible BSs deployment cases and their respective payload allowances are tabulated in Table~\ref{tab:synopsis}.
\begin{table}[!h]
\vspace{-3mm}
\centering
\caption{Different streams of Power consumption in ABSs.}
\vspace{-2mm}
\label{tab:power}
\begin{tabular}{|
>{\columncolor[HTML]{FFFFFF}}l ||
>{\columncolor[HTML]{EBF1DE}}l |
>{\columncolor[HTML]{FDEADA}}l |
>{\columncolor[HTML]{FCFBBC}}l |}
\hline
\textbf{ABS}        & $P_{T,(.)}$ [kW] & {\color[HTML]{000000} \textbf{\%P$_{w/o BS}$}} & \textbf{\%P$_{Comb}$}   \\ \hline\hline
\textbf{RWD-PiC}  & \textbf{2.04}     & {\color[HTML]{000000} \textbf{74\%}}      & \textbf{26\%}   \\ \hline
\textbf{RWD-MiC} & \textbf{4.04}     & {\color[HTML]{000000} \textbf{37\%}}      & \textbf{63\%}   \\ \hline \hline
\textbf{FWD-MiC (5 m)}   & \textbf{1.15}       & {\color[HTML]{000000} \textbf{71\%}}      & \textbf{29\%}    \\ \hline
\textbf{FWD-MiC (10 m)}   & \textbf{2.39}       & {\color[HTML]{000000} \textbf{91\%}}      & \textbf{9\%}    \\ \hline \hline
\textbf{HAP-MaC (30 m)}   & \textbf{1.78}       & {\color[HTML]{000000} \textbf{17\%}}      & \textbf{83\%}       \\ \hline
\textbf{HAP-MaC (60 m)}   & \textbf{2.68}       & {\color[HTML]{000000} \textbf{45\%}}      & \textbf{55\%}       \\ \hline \hline
\textbf{HAP-Split (25 m)}   & \textbf{1.45}       & {\color[HTML]{000000} \textbf{24\%}}      & \textbf{76\%}       \\ \hline
\textbf{HAP-Split (35 m)}   & \textbf{1.65}       & {\color[HTML]{000000} \textbf{33\%}}      & \textbf{67\%}       \\ \hline
\multicolumn{4}{c}{PiC: Picocell $\|$ MiC: Microcell $\|$   MaC: Macrocell $\|$   Split: Antenna only}
\end{tabular}
\vspace{-2mm}
\end{table}
\begin{table}[!h]
\centering
\caption{Feasibility and compatibility summary.}
\vspace{-2mm}
\label{tab:synopsis}
\resizebox{\columnwidth}{!}{%
\begin{tabular}{|c|c|cc|cccc|}
\hline
\textbf{ABS} & \textbf{RWD} & \multicolumn{2}{c|}{\textbf{FWD}} & \multicolumn{4}{c|}{\textbf{HAP}} \\ \hline
\textbf{\begin{tabular}[c]{@{}c@{}}Wingspan [m]\end{tabular}} & - & \multicolumn{1}{c|}{5} & 10 & \multicolumn{1}{c|}{30} & \multicolumn{1}{c|}{60} & \multicolumn{1}{c|}{25} & 35 \\ \hline
\textbf{\begin{tabular}[c]{@{}c@{}}Max. BS mass [kg]\end{tabular}} & 12 & \multicolumn{1}{c|}{11} & 10 & \multicolumn{1}{c|}{45} & \multicolumn{1}{c|}{270} & \multicolumn{1}{c|}{5} & 15 \\ \hline
\textbf{Feasible BS} & PiC, MiC & \multicolumn{2}{c|}{PiC, MiC} & \multicolumn{2}{c|}{\begin{tabular}[c]{@{}c@{}}PiC, MiC,\\ MaC\end{tabular}} & \multicolumn{2}{c|}{\begin{tabular}[c]{@{}c@{}}PiC, Split\end{tabular}} \\ \hline
\textbf{Compatible BS} & PiC, MiC & \multicolumn{2}{c|}{MiC} & \multicolumn{2}{c|}{MaC} & \multicolumn{2}{c|}{\begin{tabular}[c]{@{}c@{}}Split\end{tabular}} \\ \hline
\multicolumn{8}{c}{PiC: Picocell $\|$ MiC: Microcell $\|$   MaC: Macrocell $\|$   Split: Antenna only}
\end{tabular}%
}
\end{table}

\vspace{-3mm}
\section{Network Dimensioning and A Case Study}
\vspace{-1mm}
\label{sec:cov ana}



Our analysis in Section~\ref{sec:Serv time} analysis reveals RWD-BS deployment's vulnerability in terms of service time. As a result, we select it as a case study to evaluate network dimensioning and deployment costs for achieving efficient aerial network coverage. We assume no co-existence with the terrestrial network for a particular area of interest using RWD-BS hovering $h=100$ m above the ground. To determine the number of RWDs required to cover the area of interest, we first calculate the RWD-BS coverage radius $r_c$, which is affected by several factors such as altitude, frequency of operation of the BS, and the environment. The $r_c$ is modeled such that the cell-edge user experiences an outage probability of less than $1$\%. The total path loss (in dB)~\cite{b17} at the cell edge is given by
\vspace{-2mm}
\begin{equation}
\label{eq:PL}
    \phi=FSPL+\eta_{ex},
\vspace{-1mm}
\end{equation}
where $FSPL$ is free space path loss~\cite{b16} given in \eqref{eq:fspl}, $\eta_{ex}$ is the excess path loss due to line-of-sight (LoS) and non-LoS (NLoS) channels and is given in~\eqref{eq:exPL}
\vspace{-2mm}
\begin{equation}
\label{eq:fspl}
    FSPL=20\log d +20\log f -27.55,
\vspace{-1mm}
\end{equation}
where $d=\sqrt{h^2+r_c^2}$ is the distance from the RWD-BS hovering at altitude $h$ to the receiver on the ground at $r_c$, $f$ is the operational frequency in MHz. The excess path loss in probabilistic terms~\cite{b17} is given by 
\vspace{-2mm}
\begin{equation}
\label{eq:exPL}
    \eta_{ex}=\mathrm{P_{L}}\boldsymbol{\mathcal{P}}(\mathrm{L},\theta)+\mathrm{P_{NL}}\boldsymbol{\mathcal{P}}(\mathrm{NL},\theta),
\vspace{-1mm}
\end{equation}
where $\boldsymbol{\mathcal{P}}(\mathrm{L},\theta)$ is the probability of LoS given in~\eqref{eq:probLoS} which is a function of $\theta=(\tan^{-1}(h/r))$ and $\boldsymbol{\mathcal{P}}(\mathrm{NL},\theta)=(1-\boldsymbol{\mathcal{P}}(\mathrm{L},\theta))$ is the probability of NLoS.  $\mathrm{P_{L}}$ and $\mathrm{P_{NL}}$ are the LoS and the NLoS path loss, respectively~\cite{b16} and are given by 
\vspace{-2mm}
\begin{equation}
\label{eq:PLlos}
    \mathrm{P}_{\xi}=\mathcal{N}(\mu_{\xi},\sigma^2_{\xi}(\theta)),
\vspace{-1mm}
\end{equation}
where $\mathcal{N}$ is normal distribution with mean $\mu_{\xi}$ and standard deviation $\sigma_{\xi}(\theta)=a_{\xi}\exp(-b_{\xi}\theta)$; $\xi\in\{\mathrm{L, NL}\}$ is the propagation group indicator. $a_{\xi}$ and $b_{\xi}$ are frequency and environment dependent parameters in~\cite{b16}. The probability of LoS is calculated as follows:
\vspace{-2mm}
\begin{equation}
\label{eq:probLoS}
    \boldsymbol{\mathcal{P}}(\mathrm{L},\theta)=g_1(\theta-\theta_0)^{g_2},
\vspace{-1mm}
\end{equation}
where $g_1$ and $g_2$ are frequency and environment dependent parameters in~\cite{b16}, and $\theta_0$ is $15^\circ$ for this model setup.

Using~\eqref{eq:PL}, we calculate the received signal power as $P_{rx}=P_{tx}-\phi$, where $P_{tx}$ is the transmission power. From the above equations, using Monte-Carlo simulations, the coverage radius is calculated as the maximum radius $R_c$ at which the $P_{rx}$ is greater than the minimum receiver sensitivity ($P_{min}$) for more than $99\%$ of the time and is given by
\vspace{-2mm}
\begin{equation}
\label{eq:Radiuscoverage}
    \max_{r_c=R_c} \{\boldsymbol{\mathcal{P}}(P_{rx}>P_{min})>0.99\},
\end{equation}

To design and analyze the cost of the network, we need to know how many ABSs are needed to cover the service area and how many backup batteries are needed to support each ABS to run for an indefinite time, considering the battery hot-swapping technique~\cite{b18}. For a circular coverage radius $R_c$, the number of ABS required to cover the service area $A_{s}$ is given by $ N_{ABS}=\lceil{A_{s}/(\pi R_c^2})\rceil$. Assuming enough charging slots available at the charging station to recharge each battery after depletion, the number of backup batteries required for each ABS to serve an indefinite time is given as the ratio of charging time $T_{c}$ to service time $T_{s,R}$:
\vspace{-1mm}
\begin{equation}
    \label{eq:noofBB}
    N_{BB}=\lceil{\frac{T_{c}}{T_{s,R}}}\rceil =\lceil{\frac{E_dm_{b}/P_{c}}{E_dm_{b}/P_{T,R}}}\rceil=\lceil{\frac{P_{T,R}}{P_{c}}}\rceil,
\end{equation}
where $P_{c}$ is the charging power supply to the battery. A very important result observed in~\eqref{eq:noofBB} is that the number of backup batteries is independent of the battery capacity ($E_dm_{b}$). Furthermore, $T_c$, and thereby $N_{BB}$, decreases with increase in $P_c$. 
\vspace{-2mm}
\section{Results}
\label{sec:res}
For the simulation setup~\cite{b16}, urban environment is considered with $f=2000$ MHz in~\eqref{eq:fspl}, $\mu_\xi=\{1,20\}$ dB, $a_\xi=\{10.39,29.6\}$, $b_\xi=\{0.05,0.03\}$ in~\eqref{eq:PLlos}, $g_1=0.6$ and $g_2=0.11$ in~\eqref{eq:probLoS}. $\theta$ is varied from ($\pi/2$ to $\pi/12$) rad for $1000$ Monte-carlo simulations and $R_c$ satisfying~\eqref{eq:Radiuscoverage} is simulated. $N_{ABS}$ are calculated for $A_s=1$ km$^2$. Assuming $P_c=300$ W, $N_{BB}$ is calculated using~\eqref{eq:noofBB}. The resulting aspects of network dimensioning and deployment costs for picocell and microcell are tabulated in Table~\ref{tab:network dim}. For picocells, up to $7$-$10$ backup batteries per ABS (range is due to varying battery mass from $5$ kg to $9$ kg) are required for continuous operation, while for microcells, up to $14$-$17$ backup batteries per ABS are needed.

Consequently, serving a $1$ km$^2$ area would demand a minimum of $42$ to a maximum of $60$ backup batteries. This leads to a significant battery mass of $210$ kg to $540$ kg, requiring multiple charging stations capable of parallelly charging the $N_{BB}$ batteries. Although using RWD-BSs is impractical due to storage limitations and the infrastructure required for the charging station, it is evident that cost factors, such as the number of ABSs ($N_{ABS}$), type of BS deployment, and capacity requirements, must be carefully considered when deploying aerial networks to cover large areas.
\begin{table}[!h]
\vspace{-3mm}
\centering
\caption{Network dimensioning for the RWD to cover $1$ km$^2$ area.}
\vspace{-2mm}
\label{tab:network dim}
\begin{tabular}{|c|c|c|}
\hline
\textbf{BS}                                                                                   & \textbf{Picocell} & \textbf{Microcell} \\ \hline
\begin{tabular}[c]{@{}c@{}}$P_{tx}$ [W]\end{tabular}                       & 0.13               & 6.4                 \\ \hline
BS Mass [kg]                                                                                 & 3                  & 12                  \\ \hline
$r_c$=$R_c$ [m]                                                                                      & 251                & 351                 \\ \hline
\textbf{\begin{tabular}[c]{@{}c@{}}$N_{ABS}$\end{tabular}} [\#]                     & 6                  & 3                   \\ \hline
\textbf{\begin{tabular}[c]{@{}c@{}}$N_{BB}$\end{tabular}} [\#]     & 7-10               & 14-17               \\ \hline
\textbf{\begin{tabular}[c]{@{}c@{}}$N_{ABS}*N_{BB}$\end{tabular}} [\#] & 42-60              & 42-51               \\ \hline
\end{tabular}
\end{table}
\vspace{-4mm}
\section{Conclusions}
\vspace{-1mm}
\label{sec:conclu}

In this paper, we presented a comprehensive analysis of different ABSs including RWDs, FWDs, and HAPs. Key performance metrics such as power consumption, energy harvested-to-consumption ratio, and service time are investigated for different ABS configurations, considering factors such as wingspan, battery capacity, and geographical region. Our analysis showed that the FWDs have longer service times when compared to RWD-BS and HAPs have energy harvested-to-consumption ratios greater than one, indicating theoretically indefinite service time. For RWDs, solar harvesting does not significantly improve the service time, leading to the conclusion that not installing solar panels is more practical. It is evident from the energy harvesting-to-consumption ratio that solar harvesting can extend the service time for FWDs and can achieve indefinite service times for HAPs due to their larger wing area. Some small-cell solutions (pico, micro) may be compatible with RWDs and FWDs, while conventional BSs for macrocells are typically too heavy for these aerial platforms. Furthermore, conventional BSs may not be certified for stratospheric operations. The cost-benefit analysis of network dimensioning analysis demonstrates that RWD-BS deployment is impractical for achieving long-term coverage due to the substantial number of batteries required and the infrastructure needed for charging stations. Therefore, alternative ABS configurations such as FWD-BS and HAP-BS may be considered for achieving prolonged and uninterrupted service. 

\vspace{-1mm}
\section*{Acknowledgement}
\vspace{-2mm}
This work was supported in part by the CELTIC-NEXT Project, 6G for Connected Sky (6G-SKY), with funding received from Vinnova, Swedish Innovation Agency, and the Federal Ministry for Economic Affairs and Climate Action under the contract number 01MJ22010B. The views expressed herein can in no way be taken to reflect the official opinion of the German ministry.


\vspace{-1mm}
\begin{thebibliography}{00}
    \bibitem{b1} Y. Huo \emph{et al.}, ``Distributed and Multilayer UAV Networks for Next-Generation Wireless Communication and Power Transfer: A Feasibility Study," \emph{IEEE Internet Things J.,} vol. 6, no. 4, pp. 7103-7115, Aug. 2019.
    
    \bibitem{b9} S. C. Arum \emph{et al.}, ``Energy management of solar-powered aircraft-based high altitude platform for wireless communications," \emph{MDPI Electron.,} vol. 9, no. 1, pp. 1-25, 2020.

    \bibitem{bb1} H. Jin \emph{et al.}, ``A survey of energy efficient methods for UAV communication," \emph{Veh. Commun.,} p.100594, 2023.

    \bibitem{bb2} X. Jiang \emph{et al.},  ``Green UAV communications for 6G: A survey," \emph{Chin. J. Aeronaut.,} pp.19-34, 2022.

    \bibitem{b3} C. Desset \emph{et al.}, ``Flexible power modeling of LTE base stations," in \emph{Proc. IEEE Wireless commun. Netw. Conf.
    (WCNC),} Apr. 2012, pp. 2858-2862. 

    \bibitem{b12} G. Auer \emph{et al.}, ``How much energy is needed to run a wireless network?," \emph{IEEE 
 Wireless Commun.,} vol. 18, no. 5, pp. 40-49, 2011.
    
    \bibitem{b8} M. Deruyck, W. Joseph,  and L. Martens, ``Power consumption model for macrocell and microcell base stations," \emph{Trans. Emerg. Telecommun. Technol.,} vol. 25, no. 3, pp.320-333, 2014.

    \bibitem{b13} J. Yang, Q. Bao, L. Shen, and L. Ding, ``Potential applications for perovskite solar cells in space," \emph{Nano Energy,} vol. 76, pp. 105019, 2020.

    \bibitem{b14} 
    I. Ridwana and B. Alfindo, ``The effect of use of solar panels on micro scale fixed-wing UAV type as a power recharging system," in \emph{AIP Conf. Proc.}, 2019, pp. 1-8. 

    \bibitem{b15} M. Bronz \emph{et al.}, ``Towards a long endurance MAV," \emph{Int. J. Micro Air Veh.,} vol. 1, no. 4, pp. 241-254, 2009.
    
    \bibitem{b10} Y. Zeng, J. Xu, and R. Zhang, ``Energy minimization for wireless communication with rotary-wing UAV," \emph{IEEE Trans. Wireless Commun.,} vol. 18, no. 4, pp. 2329-2345, 2019.

    \bibitem{b11} Y. Zeng and R. Zhang, ``Energy-efficient UAV communication with trajectory optimization," \emph{IEEE Trans. Wireless Commun.,} vol. 16, no. 6, pp. 3747-3760, 2017.

    \bibitem{a1} P. S. Ramesh and J. M. L. Jeyan, ``Comparative Analysis of Fixed-Wing, Rotary-Wing and Hybrid Mini Unmanned Aircraft Systems (UAS) from the Applications Perspective," \emph{INCAS Bulletin,} vol. 14, no. 1, pp. 137-151, 2022.

    \bibitem{b18} S. A. H. Mohsan \emph{et al.}, ``A comprehensive review of micro UAV charging techniques," \emph{Micromachines,} vol. 13, no. 6, pp. 1-30, 2022.

    \bibitem{b17} A.Al-Hourani \emph{et al.}, ``Optimal LAP altitude for maximum coverage," \emph{IEEE Wireless Commun. Lett.,} vol. 3, no. 6, pp. 569-572, 2014.

    \bibitem{b16} A. Al-Hourani \emph{et al.}, ``Modeling air-to-ground path loss for low altitude platforms in urban environments," in \emph{Proc. IEEE GLOBECOM,} 2014, pp. 2898-2904.


    



    


    
\end{thebibliography}
\end{document}